\documentclass[conference]{IEEEtran}
\usepackage{amsmath,amssymb,dsfont,stfloats,color,url,mathtools}
\usepackage{tikz}
\usepackage{comment}
\usetikzlibrary{shapes.geometric, arrows}
\tikzstyle{startstop} = [rectangle, rounded corners, minimum width=3cm, minimum height=1cm,text centered, draw=black, fill=red!30]
\tikzstyle{process} = [rectangle, minimum width=3cm, minimum height=1cm, text centered, draw=black, fill=blue!30]
\tikzstyle{decision} = [diamond, minimum width=3cm, minimum height=1cm, text centered, draw=black, fill=green!30]
\tikzstyle{arrow} = [thick,->,>=stealth]
\usepackage[]{graphicx}
\usepackage{wasysym,esint}
\usepackage{cite}
\usepackage{caption}
\usepackage{subcaption}
\usepackage{gensymb} 
\usepackage[boxruled,linesnumbered]{algorithm2e}
\usepackage{algorithm2e}
\usepackage{lipsum}

\DeclareGraphicsExtensions{.eps,.pdf,.png,.jpg,.gif,.jpeg,.pstex}
\newtheorem{rem}{Remark}

\newcommand{\twodots}{\mathinner {\ldotp \ldotp}}
\newfont{\bbb}{msbm10 scaled 700}

\newfont{\bb}{msbm10 scaled 1100}
\newcommand{\CC}{\mbox{\bb C}}

\newcommand{\RR}{\mbox{\bb R}}

\newcommand{\av}{{\bf a}}

\newcommand{\uv}{{\bf u}}
\newcommand{\wv}{{\bf w}}
\newcommand{\vv}{{\bf v}}

\newcommand{\Sm}{{\bf S}}
\newcommand{\Tm}{{\bf T}}
\newcommand{\Um}{{\bf U}}
\newcommand{\Wm}{{\bf W}}
\newcommand{\Vm}{{\bf V}}

\newcommand{\Lc}{{\cal L}}

\newcommand{\herm}{{\sf H}}

\newcommand{\transp}{{\sf T}}

\usepackage{stmaryrd} % for \mkv 

\usepackage{hyperref}
\hypersetup{
    bookmarks=true,         % show bookmarks bar?
    unicode=false,          % non-Latin characters in AcrobatÕs bookmarks
    pdftoolbar=true,        % show AcrobatÕs toolbar?
    pdfmenubar=true,        % show AcrobatÕs menu?
    pdffitwindow=false,     % window fit to page when opened
    pdfstartview={FitH},    % fits the width of the page to the window
    pdfnewwindow=true,      % links in new window
    colorlinks=true,       % false: boxed links; true: colored links
    linkcolor=red,          % color of internal links (change box color with linkbordercolor)
    citecolor=green,        % color of links to bibliography
    filecolor=blue,      % color of file links
    urlcolor=blue           % color of external links
}
\title{Flat-Top Beamforming with Efficient Array-Fed RIS}

\author{\IEEEauthorblockN{Krishan K. Tiwari, Giuseppe Caire}
	\IEEEauthorblockA{Technische Universit\"at Berlin, Germany}
	 {lastname}@tu-berlin.de
}

\begin{document}

\bstctlcite{BSTcontrol} 

%\date{10 November 2021}
\maketitle

\begin{abstract}
Flat-top beam designs are essential for uniform power distribution over a wide angular sector for applications such as 5G/6G networks, beaconing, satellite communications, radar systems, etc. Low sidelobe levels with steep transitions allow negligible cross sector illumination. Active array designs requiring amplitude taper suffer from poor power amplifier utilization. Phase only designs, e.g., Zadoff-Chu or generalized step chirp polyphase sequence methods, often require large active antenna arrays which in turns increases the hardware complexity and reduces the energy efficiency. In our recently proposed novel array-fed reflective intelligent surface (RIS) architecture, the small ($2 \times 2$) active array has uniform (principal eigenmode) amplitude weighting. We now present a pragmatic flat-top pattern design method for practical array (RIS) sizes, which outperforms current state-of-the-art in terms of design superiority, energy efficiency, and deployment feasibility. 
This novel design holds promise for advancing sustainable wireless technologies in next-generation communication systems, including applications such as beaconing, broadcast signaling, and hierarchical beamforming, while mitigating the environmental impact of high-energy antenna arrays.
\end{abstract} 

%%%%%%%%%%%%%%%%%%%%%%%%%%%%%%%%%%%%%%%%%%%%%%%%

%\vspace{-0.5cm}
\section{Introduction}  
\label{sec:intro}

\IEEEPARstart{F}{lat} top power radiation patterns from antenna arrays have been a subject of intense research for several decades. Finite impulse response (FIR) filter based techniques for active arrays \cite{firpm} require amplitude taper across the active array, which in turn means poor power amplifier (PA) utilization and lower energy efficiency. Therefore, phase-only techniques have been proposed \cite{GSC} which work well for large active arrays, in the sense that the passband ripple is not very attractive for small to medium array sizes, e.g., we get about 6-7 dB ripple for $40$ elements linear and 14 dB for $40 \times 40 ~(= 1600)$ planar configurations! However, very large active arrays are not very energy efficient due to the fact that the PA output power of individual antenna elements becomes very low, requiring the use of power splitting networks with their own insertion losses! 

\begin{figure}[t!]
\centerline{\includegraphics[width=8.25cm]{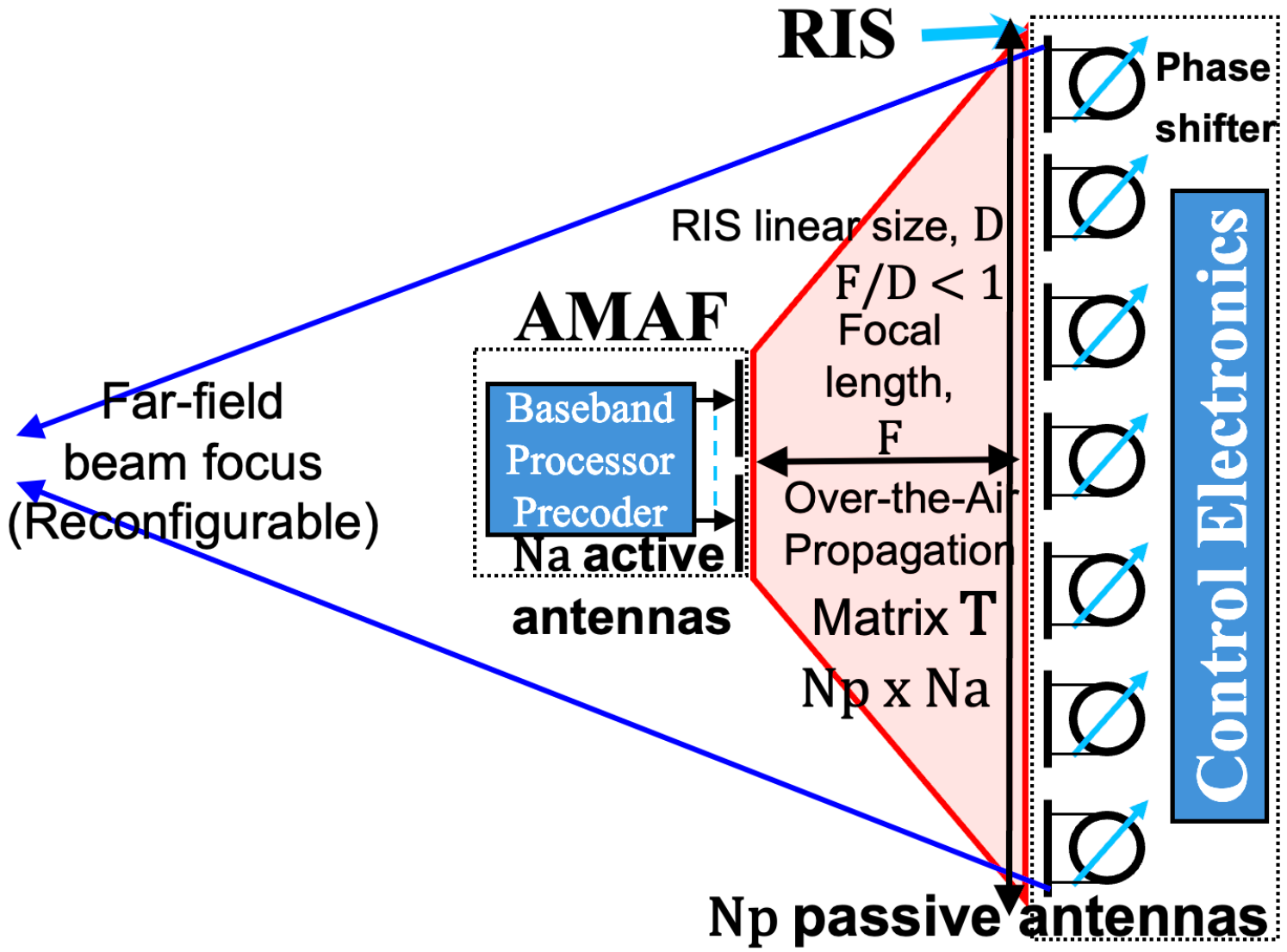}}
	\caption{RIS fed by an AMAF placed in its near field. $F/D < 1$.}
	\label{fig:sysmo}
\end{figure}

The energy efficiency would be the best if we could use uniformly weighted small active arrays (for good PA utilization) and at the same time the enhanced gain, degrees-of-freedom, and agile performance capabilities of larger antenna arrays. In this direction, we had proposed passive (reflect/transmit)arrays (a.k.a. reflective intelligent surfaces (RIS)) fed by a small active multi-antenna feeder (AMAF) \cite{VTC2022, ICC2023, VTC2024, SPAWC2024}, see Fig. \ref{fig:sysmo}. 
The AMAF-RIS matrix $\Tm$ (see Fig. \ref{fig:sysmo} and \eqref{eq:T}) 
can be seen as a transformer between the small active array and the large passive RIS, effectively combining the benefits of both the systems.
Recall that due to the inherent symmetry of the center feed geometry, the $2 \times 2$ AMAF principal eigenmode will always be the (practically) uniform weighting for maximum AMAF-RIS power transfer, resulting in the best PA utilization. For simplicity, we consider Friis-modeling based AMAF-RIS matrix $\Tm \in \CC^{N_p \times N_a}$ given by
\begin{equation}
\label{eq:T}
\centering
T_{n,m} = \frac{\sqrt{E_A(\theta_{n,m})E_R(\phi_{n,m}})~ \exp(-j\pi r_{n,m})}{2\pi r_{n,m}},
\end{equation}
where $E_A, ~E_R, ~\theta_{n,m},~ \phi_{n,m}, ~ \text{and}~r_{n,m}$ are AMAF element power radiation pattern, RIS element power radiation pattern, angle of departure from an AMAF element $m$ to a RIS element $n$ measured from the AMAF element boresight, angle of arrival from the AMAF element $m$ to the RIS element $n$ measured from the RIS element boresight, and the distance (normalized by $\lambda/2$) between the AMAF element $m$ and the RIS element $n$, respectively, and we let $j:=\sqrt{-1}$. In \cite{FWValidation}, we had validated this simple approach by an accurate full-wave solution which inherently takes into account inter-element mutual coupling, polarization, and element antenna phase pattern effects.

%\vspace{-12pt}

The synthesis of flat top beams from such an AMAF-RIS architecture is a very difficult problem due to the following constraints: i) the passive RIS elements have only phase control and no amplitude control, ii) the principal eigenmode (PEM) profile at the RIS (both amplitude and phase) depends on the F/D ratio, is obtained from the singular value decomposition (SVD) of the AMAF-RIS propagation matrix $\Tm$, and therefore cannot be represented in closed form, iii) the PEM amplitude taper \cite[Fig. 2]{ICC2023} makes it difficult to apply algebraic phase-only techniques \cite{GSC} for constant modulus polyphase sequences, and iv) from a mathematical optimization vantage point, it's a very high-dimensional, non-convex, non-trivial problem with a very low probability of success, in the sense that useless local optima are far more numerous than useful local optima!

We now present in Section \ref{sec:flat_top_beam_design} a pragmatic method for getting the RIS phases for obtaining flat-top beams from AMAF-RIS principal eigenmode\footnote{Note that any joint AMAF-RIS optimization which deviates from the PEM leads to a loss in the AMAF-RIS power transfer. Therefore, the PEM is recommended to maintain the maximum AMAF-RIS power transfer.}, which are a good starting point for further improved flat-top beams using mathematical optimization. In Section \ref{sec:examples}, we present full 3D design examples and ground footprints of beams so obtained for (pico)cellular applications. Section \ref{sec:ee_comp} provides the energy efficiency comparison against active arrays. Section \ref{sec:CONC} concludes the paper.

%{\bf Mathematical Notations:} $\mathbb{R}$ and $\mathbb{C}$ denote the sets of all real and all complex numbers, respectively. $\text{x}^*$ denotes conjugate of a complex scalar, $\|\xv\|$ is the Euclidean norm of a vector $\xv$, $[\cdot]^{\circ2}$ denotes Hadamard (elementwise) square, $[\cdot]^\transp$ is transpose, $[\cdot]^\herm$ is the Hermitian transpose, $|\xv|$ and $\angle \xv$ denote vectors containing the magnitudes and the angles (phases) of the elements in $\xv$, respectively. $|\Xm|$ is a matrix containing the magnitudes of the elements in $\Xm$, and $\text{diag}(\xv)$ denotes a square diagonal matrix whose diagonal elements are given by $\xv$. $\lceil . \rceil$ denotes the ceiling function. $j=\sqrt{-1}.$ %All lengths/distances are normalized by $\lambda/2$, unless specified otherwise.
%%%%%%%%%%%%%%%%%%%%%%%%%%%%%%%%%%%%%%%%%%%%%%%%

\section{Pragmatic Design of Flat-Top Beams} 
\label{sec:flat_top_beam_design}

In order to mitigate the difficulty due to the high-dimension, non-convex nature of the problem for planar arrays, we obtain the phase-only beamforming weights for standard linear arrays (SLAs) and then use the straightforward outer product for planar array designs, and thus recommend to tackle the problem in the lower-dimensional space of linear arrays instead of dealing directly with the complexity of planar arrays as in \cite{ghanem2022optimization}. 

The proposed sequential method has the following steps:
\begin{enumerate}
    \item For linear AMAF ($N_a=2$) and RIS ($N_p=40$) arrays, choose the F/D ratio such that the PEM magnitude taper is suitable, e.g., matches well to the envelope/main lobe of
the amplitude profile obtained from a flat-top beam shape obtained from ``firpm'' (a MATLAB\textregistered\ function utilizing \cite{firpm}) based design for active arrays. For example,
F/D = 0.235 as we will see in the next section.
   \item Evenly about the RIS center, choose suitable RIS elements groupings for binary phases ($0$ or $\pi$ rad), e.g., nearest integer to 0.15 to 0.18 of the linear size $N_p$. We denote this vector as $\wv_{\rm binary}$. The binary phase sequence $\wv_{\rm binary}$ applied to the PEM amplitude profile will yield a flat-top beam. Mathematically, this can be represented as
\begin{equation} \label {eq:binary_PEM}
  \wv_{\rm binary} \odot \widetilde{\wv} \odot \uv_1,
\end{equation}
where $\widetilde{\wv}=\left [ \exp(-j \angle \uv_1) \right ]$ is a vector of phase shifts such that $|\uv_1|=\widetilde{\wv} \odot \uv_1 \in \RR_+^{N_p}$, $\odot$ is Hadamard (elementwise) product, $\angle \uv_1$ is the vector of phases of $\uv_1$, $\exp$ is applied componentwise, 
and $\uv_1$ is first column vector of $\Um \in \CC^{N_p \times N_p}$ obtained from the SVD of $\Tm=\Um\Sm\Vm^\herm$.
     
   In general, the width of the so obtained flat-top beam is not widely tunable because:
   a) the RIS amplitude taper is fixed for a given F/D ratio, and b) an expansion in one Fourier domain is compression in another Fourier domain and vice versa. 
   \item Use the beam widening function \cite{R1-1611929, R1-1700772} or equivalently the phase perturbation function (PPF)
\begin{equation} \label{eq:phase_perturbation_function}
    f(n) = \left| 4 \pi c \left( \frac{0.5}{N_p-1} + \frac{n - 0.5N_p}{N_p-1} \right)^p \right|,
\end{equation}
where $c$, $N_p$, and $n \in [0,N_p-1]$ are the scaling parameter for the phase perturbation, the number of antenna elements, and the antenna element index, respectively. The parameter $p$ controls the exponent power applied to the position-dependent term. The beam broadening vector $\wv_{\rm PPF}$ is then
\begin{equation} \label{eq:widening_phases}
w_{\rm PPF}(n) = \exp(jf(n)). %\exp (jf(n)). e^{j f(n)}
\end{equation}
In \eqref{eq:phase_perturbation_function}, increasing $c$ results in a broader beam for a fixed $p$, whereas decreasing $p$ widens the beam for a fixed $c$.
Thus, the linear RIS phase configuration (for a template flat-top beam at the boresight) is obtained as 
\begin{equation} \label {eq:linear_RIS_phases}
 \wv = \wv_{\rm PPF} \odot \wv_{\rm binary} \odot \widetilde{\wv}.
\end{equation}
%where $\widetilde{\wv}=\left [ \exp(-j \angle \uv_1) \right ]$ is a vector of phase shifts such that $|\uv_1|=\widetilde{\wv} \odot \uv_1 \in \RR_+^{N_p}$, $\odot$ is Hadamard (elementwise) product, $\angle \uv_1$ is the vector of phases of $\uv_1$, $\exp$ is applied componentwise, 
%and $\uv_1$ is first column vector of $\Um \in \CC^{N_p \times N_p}$ obtained from the SVD of $\Tm=\Um\Sm\Vm^\herm$.% where $\Um \in \CC^{N_p \times N_p}$ and $\Vm \in \CC^{N_a \times N_a}$ are unitary matrices and $\Sm \in \CC^{N_p \times N_a}$ is a diagonal matrix containing ordered singular values $\sigma_1, \sigma_2, \twodots, \sigma_{N_a}$. 

Steps 1 to 3 provide an indispensable sequence of steps to obtain flat-top beam shapes with steep main lobe transition and low sidelobes. Furthermore, $\wv$ so obtained serves as a good starting point for the mathematical optimization algorithm in \cite[Algorithm 1]{ghanem2022optimization}. Any violation of these steps 1-3 does not seem to work.

\item The optimization algorithm in \cite{ghanem2022optimization} can be used to obtain optimized phase-only vectors $\wv_{\rm opt}$ for further improved beam shapes. In particular, it is possible to tune the flat-top beam width by specifying the optimization grid boundary. The only difference in our case from \cite{ghanem2022optimization} is that we consider the linear arrays and the array gain takes into account the principal eigenmode taper $|\uv_1|$, which is the vector containing the magnitudes of the elements of $\uv_1$ as seen in Fig. \ref{fig:step1}. Note that the given modulus constraint $|\uv_1|$ (principal eigenmode) makes the problem more difficult and non-trivial compared to constant modulus problems! For example, there is no specific restriction on the initial starting point for constant modulus problems as evident from the random initialization in \cite[Algorithm 1, step 1]{ghanem2022optimization}. It is important to note that random initialization in a high-dimensional non-convex problem is generally not expected to lead to meaningful solutions. In a non-convex cost function, if any arbitrary initialization consistently converges to a useful local optimum, it would imply either that the function is effectively convex (i.e., the local optimum is also a global optimum) or that all local optima have equivalent performance. However, in a high-dimensional non-convex setting, the probability of all local optima yielding identical objective values is vanishingly small.

\end{enumerate}

\section{Design Examples} 
\label{sec:examples}

\subsection{Linear Arrays}

\begin{figure}[h!]
\centerline{\includegraphics[width=8cm]{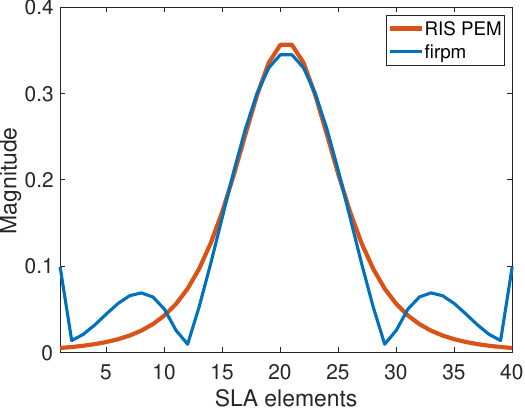}}
	\caption{Matching RIS PEM magnitude  profile to that from ``firpm'' FIR filter for a flat top beam shape.}
	\label{fig:step1}
\end{figure}

Step 1: We consider standard linear array RIS with $N_p=40$ elements fed by a standard linear array AMAF with $N_a = 2$ elements at the focal length $F=9.4$. We consider microstrip patch elements with 6 dBi gain and a half power beam width of $90 \degree$, modeled by classical axisymmetric cosine pattern, (e.g., see \cite[(14)]{jamali2020intelligent}, \cite[(17)]{Pozar_1997}, \cite[(2-31)]{Balanis_antenna_theo}),  $G_{\rm patch}(\psi) = 4 \cos^2(\psi)$,  
where $\psi$ is the angle with respect to the patch boresight. The resulting RIS magnitude profile, $|\uv_1|$, is shown in Fig. \ref{fig:step1}.

\begin{comment}
   \begin{figure*}[h!] \vspace{-0.3cm}
    \centering
 \begin{subfigure}[b]{0.428\textwidth}
        \includegraphics[width=\textwidth]{Step_2_Step_3.pdf}
        \caption{Flat top beams obtained from steps 2 and 3.}
        \label{fig:step2}
    \end{subfigure}
    \hfill
    \begin{subfigure}[b]{0.451\textwidth}
        \includegraphics[width=\textwidth]{Post_Opt_Wide_Narrow.pdf}
        \caption{Tunable width flat top beams obtained from optimization.}
        \label{fig:step4}
    \end{subfigure}
    \caption{Comparison of beam patterns: (a) beams from steps 2 and 3, and (b) tunable width flat-top beams from step 4.}
    \label{fig:combined}
\end{figure*} 
\end{comment}

\begin{figure}[h!] 
    \centering
    \vspace{-0.3cm}
    \begin{subfigure}[b]{0.428\textwidth}
        \centering
        \includegraphics[width=\textwidth]{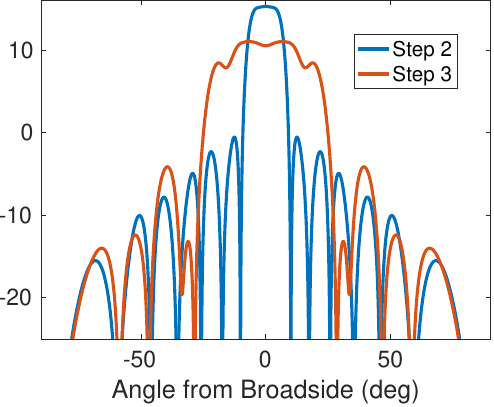}
        \caption{Flat top beams obtained from steps 2 and 3.}
        \label{fig:step2}
    \end{subfigure}
    
    \vspace{0.3cm} % Adjust spacing between figures

    \begin{subfigure}[b]{0.451\textwidth}
        \centering
        \includegraphics[width=\textwidth]{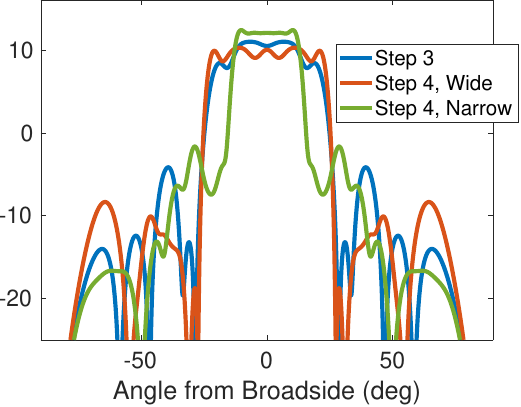}
        \caption{Tunable width flat top beams obtained from optimization.}
        \label{fig:step4}
    \end{subfigure}

    \caption{Comparison of beam patterns: (a) beams from steps 2 and 3, and (b) tunable width flat top beams from step 4.}
    \label{fig:combined}
\end{figure}

Step 2: Choose suitable groupings of RIS elements for binary phases, e.g., in this example the grouping of 7 RIS elements with $\wv_{\rm binary}= [1,1,1,1,1,1,-1,-1,-1,-1,-1,-1,-1,1,1,1,1,1,1,1,\\
1,1,1,1,1,1,1,-1,-1,-1,-1,-1,-1,-1,1,1,1,1,1,1]^\transp$ has been empirically chosen to obtain the beam shape shown in Fig. \ref{fig:step2}. These patterns are standard array theory and pattern multiplication based power radiation pattern plots given by $\big|\av^\herm(\vartheta)\wv\big|^2 E(\vartheta)$, where $E(\vartheta)$ is the element factor, $\vartheta$ is the angle from the array broadside, $\wv$ is the array beamforming vector, and $\av(\vartheta)=\left[1,e^{-j \pi \text{sin}(\vartheta)},\twodots,e^{-j \pi (N_p-1)\text{sin}(\vartheta)}\right]^\herm$ is the array steering vector, and $[\cdot]^\herm$ denotes the Hermitian transpose. For example, in this step $\wv=\wv_{\rm binary} \odot|\uv_1|$. 

%\begin{figure}[h!]
%\centerline{\includegraphics[width=8cm]{Step_2_Step_3.pdf}}
%	\caption{Flat top beams obtained from steps 2 and 3.}
%	\label{fig:step2}
%\end{figure}

Step 3: Apply the PPF obtained from \eqref{eq:phase_perturbation_function} and \eqref{eq:widening_phases} with $c=2$ and $p=1$ to obtain the widened flat top beam of Fig. \ref{fig:step2}.

Step 4: Use the optimization algorithm of \cite{ghanem2022optimization} to obtain tunable width flat-top beams as shown in Fig. \ref{fig:step4}, where we see that the passband ripple is only 1.2 dB and 0.3 dB for the wider and narrower beam shapes, respectively. Such results are not generally obtainable from \cite{GSC} because the array length tends to infinity for smooth (continuous) passbands for Zadoff-Chu or generalized step-chirp sequences. We also get better sidelobes than \cite{GSC}. 

%\begin{figure}[h!]
%\centerline{\includegraphics[width=8cm]{Post_Opt_Wide_Narrow.pdf}}
%	\caption{Tunable width flat top beams obtained from optimization with the initialization from step 3 result, the wider and the narrower beams are optimized for $+/- 22.5\degree$ and $+/-11.5\degree$, respectively.}
%	\label{fig:step4}
%\end{figure}

\begin{rem} \label{rem1}
When the desired flat-top sector is discretized with 5 grid points, we do not get a useful result even though the optimization algorithm converges. However, when the sector is discretized with 15 grid points, we get the useful results, shown above, in just three iterations.  \hfill $\lozenge$
\end{rem}

\subsection{Planar Arrays and Ground Footprints}

\begin{comment}
\begin{figure*}[ht!] \vspace{0cm}
    \centering
    \begin{subfigure}[b]{0.32\textwidth}
        \includegraphics[width=5cm]{No_Opt_Widening.pdf}
        \caption{Confined flat top beam from Step 2}
        \label{fig:confined_step_2}
    \end{subfigure}
    \hfill
    \begin{subfigure}[b]{0.32\textwidth}
        \includegraphics[width=5cm]{Opt_Widening_Az.pdf}
        \caption{Az. widened flat top beam from Step 4}
        \label{fig:azimuth_step_4}
    \end{subfigure}
    \hfill
    \begin{subfigure}[b]{0.32\textwidth}
        \includegraphics[width=5cm]{Opt_Widening_El.pdf}
        \caption{El. widened flat top beam from Step 4}
        \label{fig:elevation_step_4}
    \end{subfigure}
    \caption{Ground footprints of flat-top beams from steps 2 and 4}
    \label{fig:gnd_flat_tops}
\vspace{0cm} \end{figure*}
\end{comment}

The phase values for the planar $40 \times 40$ RIS fed by $2 \times 2$ AMAF at the $F=9.4$ is obtained by the straightforward outer product of any desired combinations of $\wv_{\rm opt}$, $\wv_{\rm PPF}\odot\wv_{\rm binary}\odot \widetilde{\wv}$, or $\wv_{\rm binary}\odot \widetilde{\wv}$, e.g., if we want to widen the flat-top beam in both azimuth and elevation, only in the azimuth, or only in the elevation.

\begin{figure}[h!]
    \centering
    \includegraphics[width=0.92\columnwidth]{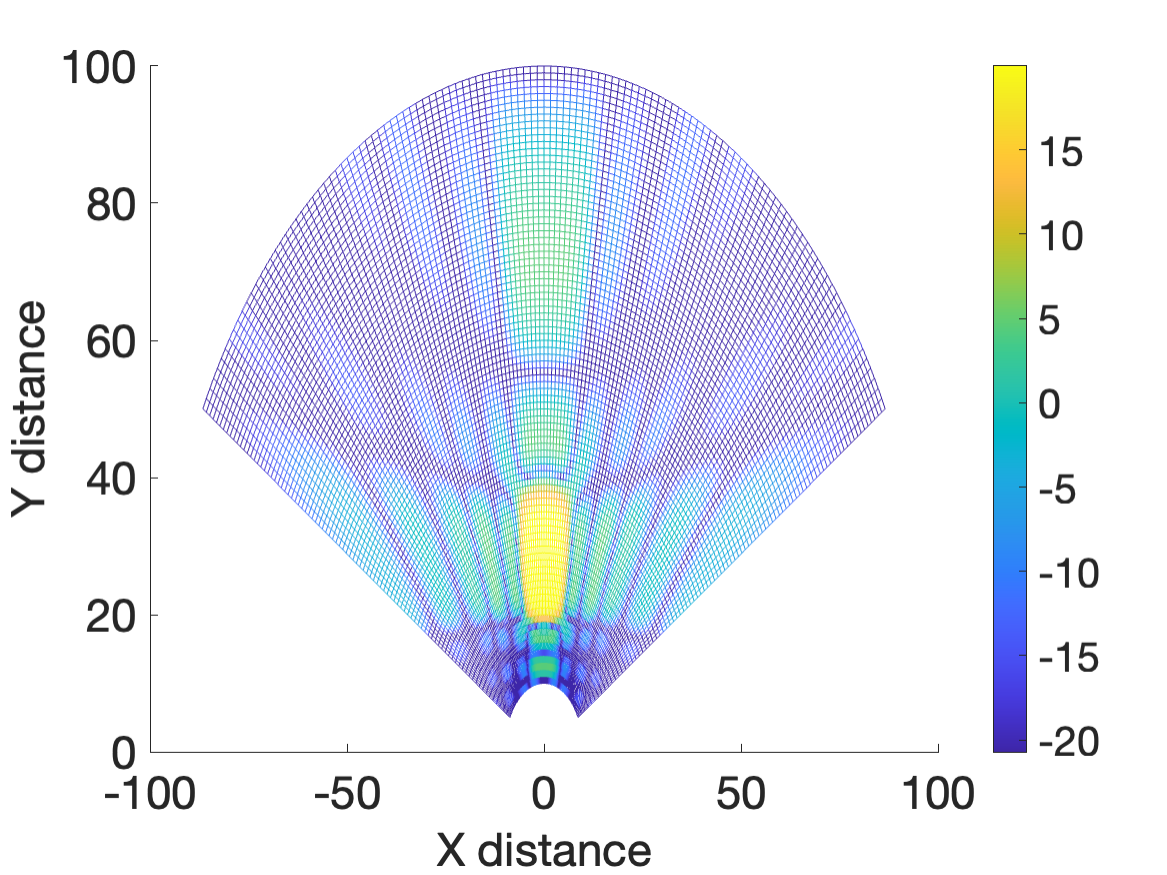}
    \caption{Confined flat top beam from Step 2.}
    \label{fig:confined_step_2}
\end{figure}

\begin{figure}[h!]
    \centering
    \includegraphics[width=0.92\columnwidth]{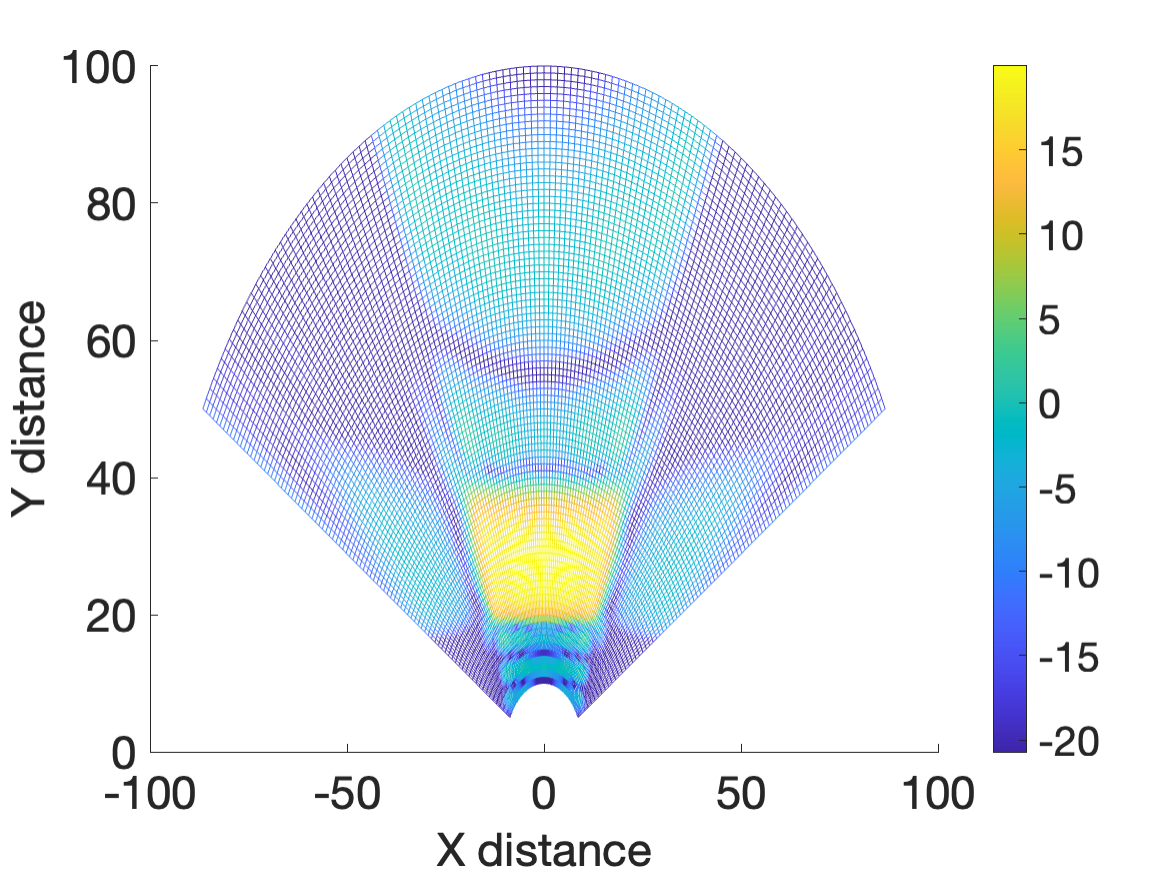}
    \caption{Azimuth widened flat top beam from Step 4.}
    \label{fig:azimuth_step_4}
\end{figure}

\begin{figure}[h!]
    \centering
    \includegraphics[width=0.92\columnwidth]{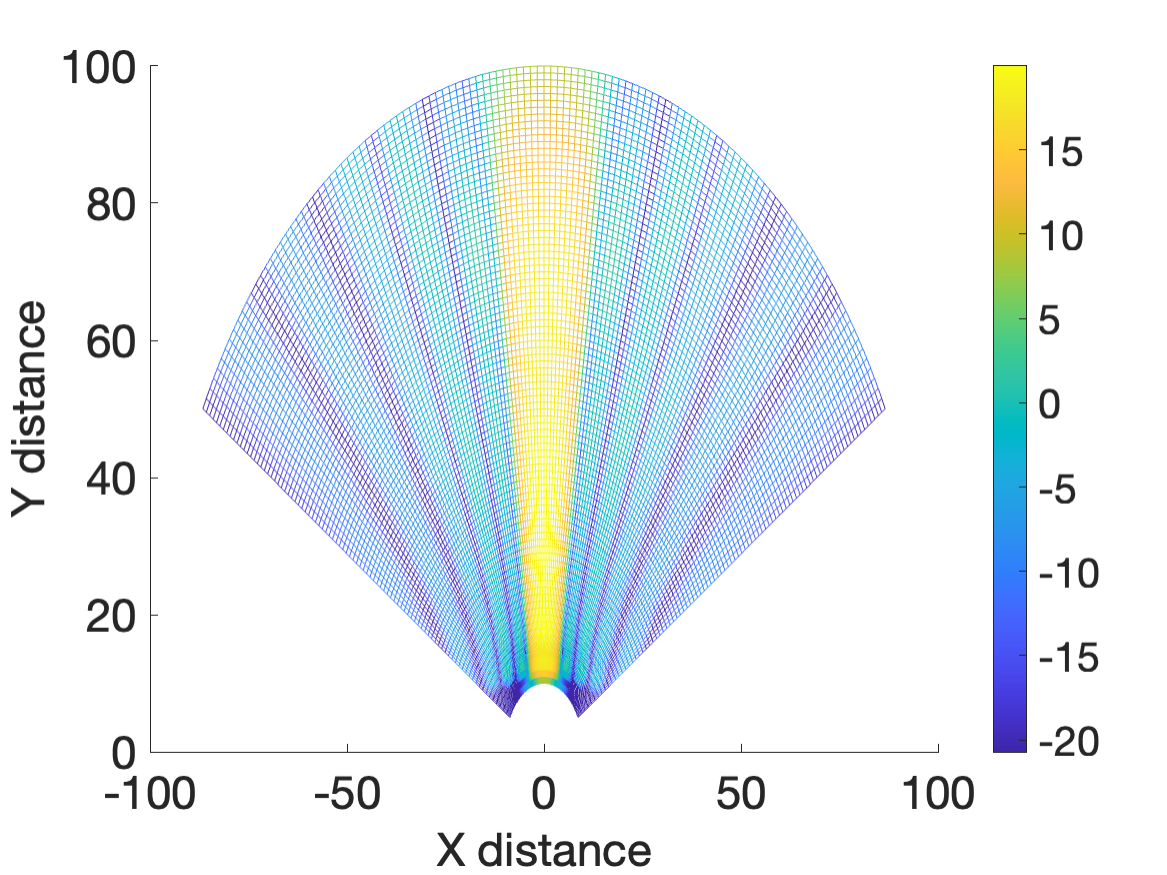}
    \caption{Elevation widened flat top beam from Step 4.}
    \label{fig:elevation_step_4}
\end{figure}

Denoting the azimuth and the elevation beamforming vectors as $\wv_{\rm az}$ and $\wv_{\rm el}$, the full planar RIS beamforming weight matrix is $\Wm = \wv_{\rm el} \wv_{\rm az}^\herm$. Using the picocell specifications of \cite[Section IV]{SPAWC2024} with the only difference of RIS size $40 \times 40$ and the focal length $F=9.4$,
%Fig. \ref{fig:gnd_flat_tops} shows 
the ground footprints of flat-top beams obtained from the phase sequences $(\wv_{\rm binary}\odot \widetilde{\wv})(\wv_{\rm binary}\odot \widetilde{\wv})^\herm$ , $(\wv_{\rm binary}\odot \widetilde{\wv}) \wv_{\rm opt}^\herm$, and $\wv_{\rm opt} (\wv_{\rm binary}\odot \widetilde{\wv})^\herm$ are shown in Figs. \ref{fig:confined_step_2}, \ref{fig:azimuth_step_4}, and \ref{fig:elevation_step_4}, respectively, on the next page. Such planar beamforming configurations are highly suitable for beaconing in picocells or indoor spaces, where uniform angular coverage is critical.

\section{Energy Efficiency Comparison}  
\label{sec:ee_comp}

We now present an energy efficiency comparison for the case where we obtain the same flat-top beam gain from an active array with the constant modulus constraint \cite{GSC} and from the proposed AMAF-RIS configuration with the given modulus constraint of the principal eigenmode (i.e., $|\uv_1|$ which is the vector containing the magnitudes of the elements of $\uv_1$ as seen in Fig. \ref{fig:step1}). For the running example from \cite[Section IV]{SPAWC2024}, let us take the RF power of $P_{\rm RF}=20~{\rm dBm}$.

For the AMAF-RIS architecture, $\vv_1=[0.5, 0.5, 0.5,0.5]$. Hence, the maximum AMAF PA output power  
is given by $P^{(1)}_{\rm pa-max}={\rm max}|v_{1i}|^2P_{\rm RF}=-6~\text{dB}+20~\text{dBm}=14~\text{dBm}=25.1~\text{mW}$. We assume that all the PAs in the (AMAF) array are developed in the same semiconductor technology, and are all biased with the same DC power dictated by the maximum requested RF power. Considering Indium Phosphide (InP) PAs with efficiency $\eta=0.3$ \cite{ETH_PA_Survey, PAlimits},  the AMAF-RIS DC power consumption, $P^{(1)}_{\rm DC} = N_a P^{(1)}_{\rm pa-max}/\eta = 4 \times 25.1~\text{mW}/0.3 = 335~\text{mW}$.

For the $40 \times 40$ active array with the constant modulus constraint \cite{GSC}, uniform element weight, $1/1600=-32~{\rm dB}$. Per PA output power, $P^{(2)}_{\rm pa-too-low} = 20~{\rm dBm}-~32\text{dB}=-12~\text{dBm}$, which is too low! Therefore, we consider a power splitter network with two stages of 4-way splitters, each output of which is fed to another splitter network with two stages of 10-way splitter, with each stage having a nominal insertion loss of 1 dB, i.e., the combined insertion loss of the overall four stage power split is 4 dB. 
Thus, the splitter network input-to-per-output-port power ratio, $\Lc = 4 + 32 = 36$ dB. This is driven by a single PA with requested RF power $P^{(2)}_{\rm pa}$ given by $P^{(2)}_{\rm pa}/\Lc=P^{(2)}_{\rm pa-too-low}$, resulting into $P^{(2)}_{\rm pa}=-12~\text{dBm}+36~\text{dB}=24~\text{dBm}=251.2~\text{mW}$.
It follows that the constant modulus active array DC power consumption, $P^{(2)}_{\rm DC} = P^{(2)}_{\rm pa}/\eta=251.2~\text{mW}/0.3=837~\text{mW}>335~\text{mW}=P^{(1)}_{\rm DC}$. Thus, for the same flat-top beam gain and the same link budget, the AMAF-RIS scheme is more energy efficient than constant modulus active arrays (which in turn have obviously better PA utilization and power efficiency than power tapered active arrays). Essentially, the AMAF-RIS architecture saves the power lost in the power splitter network which is placed after the PA stage in order to use available PAs due to too low a power level per antenna element in a (very) large active array. This is in addition to the obvious hardware simplicity and much less design and calibration effort for the space feeding. 

%%%%%%%%%%%%%%%%%%%%%%%%%%%%%%%%%%%%%%%%%%%%%%%%
%\vspace{-0.4cm}
\section{Conclusion}  
\label{sec:CONC}
%\vspace{-0.2cm}

We presented a pragmatic flat-top beam shaping method for the AMAF-RIS (or, equivalently, array-fed array) architecture. It achieves better beam shapes for moderate (RIS) array sizes than active arrays using polyphase sequences. The proposed method avoids the challenges associated with high-dimensional and non-convex optimization. Also, by virtue of the fact that the AMAF active array is of minimal size and uniformly weighted, the AMAF-RIS scheme is more energy efficient than constant modulus active arrays for the same system requirements, contributing to the evolution of sustainable and high-performance wireless connectivity.  

%%%%%%%%%%%%%%%%%%%%%%%%%%%%%%%%%%%%%%%%%%%%%%%%
%\vspace{-3pt}
\section*{Acknowledgment}
%\vspace{-3pt}
The work was supported by BMBF Germany in the program of ``Souverän. Digital. Vernetzt.'' Joint Project 6G-RIC (Project ID 16KISK030).

%%%%%%%%%%%%%%%%%%%%%%%%%%%%%%%%%%%%%%%%%%%%%%%%

%\vspace{-0.5cm}
\bibliographystyle{IEEEtran} 
\bibliography{WD-bibliography}

\begin{thebibliography}{10}
\providecommand{\url}[1]{#1}
\csname url@samestyle\endcsname
\providecommand{\newblock}{\relax}
\providecommand{\bibinfo}[2]{#2}
\providecommand{\BIBentrySTDinterwordspacing}{\spaceskip=0pt\relax}
\providecommand{\BIBentryALTinterwordstretchfactor}{4}
\providecommand{\BIBentryALTinterwordspacing}{\spaceskip=\fontdimen2\font plus
\BIBentryALTinterwordstretchfactor\fontdimen3\font minus \fontdimen4\font\relax}
\providecommand{\BIBforeignlanguage}[2]{{%
\expandafter\ifx\csname l@#1\endcsname\relax
\typeout{** WARNING: IEEEtran.bst: No hyphenation pattern has been}%
\typeout{** loaded for the language `#1'. Using the pattern for}%
\typeout{** the default language instead.}%
\else
\language=\csname l@#1\endcsname
\fi
#2}}
\providecommand{\BIBdecl}{\relax}
\BIBdecl

\bibitem{firpm}
L.~Rabiner \emph{et~al.}, ``{FIR Digital Filter Design Techniques Using Weighted Chebyshev Approximation},'' \emph{Proc. of the IEEE}, vol.~63, no.~4, pp. 595--610, Apr. 1975.

\bibitem{GSC}
C.~Du \emph{et~al.}, ``{Broad Beam Designs for Broadcast Channels},'' \emph{IEEE Trans. on Signal Process.}, vol.~72, pp. 3819--3833, Aug. 2024.

\bibitem{VTC2022}
K.~K. Tiwari \emph{et~al.}, ``{On the Behavior of the Near-Field Propagation Matrix between two Antenna Arrays, with Applications to RIS-Based Over-the-Air Beamforming.}'' in \emph{IEEE 95th Veh. Technol. Conf. (VTC2022-Spring)}, Jun. 2022, pp. 1--6.

\bibitem{ICC2023}
------, ``{RIS-Based Steerable Beamforming Antenna with Near-Field Eigenmode Feeder},'' in \emph{ICC 2023 - IEEE Int. Conf. on Commun.}, May 2023, pp. 1293--1299.

\bibitem{VTC2024}
------, ``{Power Transfer between Two Antenna Arrays in the Near Field},'' in \emph{IEEE 99th Veh. Technol. Conf. (VTC2024-Spring)}, June 2024, pp. 1--6.

\bibitem{SPAWC2024}
------, ``{A Modular Pragmatic Architecture for Multiuser MIMO with Array-Fed RIS},'' in \emph{SPAWC 2024 - The 25th IEEE Int. Workshop on Signal Process. Advances in Wireless Commun.}, Sep. 2024, pp. 556--560.

\bibitem{FWValidation}
------, ``{Array-Fed RIS: Validation of Friis-Based Modeling Using Full-Wave Simulations},'' in \emph{The 16th German Microwave Conf. (GeMiC), Dresden}, Mar. 2025, pp. 611--614.

\bibitem{ghanem2022optimization}
W.~R. Ghanem \emph{et~al.}, ``{Optimization-Based Phase-Shift Codebook Design for Large IRSs},'' \emph{IEEE Commun. Lett.}, vol.~27, no.~2, pp. 635--639, Feb. 2023.

\bibitem{R1-1611929}
Intel, ``Codebook with beam broadening,'' vol. {R1-1611929 in 3GPP TSG-RAN WG1 87}, Nov. 2016.

\bibitem{R1-1700772}
Ericsson, ``On forming wide beams,'' vol. {R1-1700772 in 3GPP TSG-RAN WG1 87ah-NR}, Jan. 2017.

\bibitem{jamali2020intelligent}
V.~Jamali \emph{et~al.}, ``{Intelligent Surface-Aided Transmitter Architectures for Millimeter-Wave Ultra Massive MIMO Systems},'' \emph{IEEE Open J. of the Commun. Society}, vol.~2, pp. 144--167, Dec. 2020.

\bibitem{Pozar_1997}
D.~Pozar \emph{et~al.}, ``{Design of Millimeter Wave Microstrip Reflectarrays},'' \emph{IEEE Trans. on Antennas and Propag.}, vol.~45, no.~2, pp. 287--296, Feb. 1997.

\bibitem{Balanis_antenna_theo}
C.~A. Balanis, \emph{Antenna Theory, Analysis and Design}.\hskip 1em plus 0.5em minus 0.4em\relax Wiley, 2016.

\bibitem{ETH_PA_Survey}
\BIBentryALTinterwordspacing
H.~Wang \emph{et~al.} Power amplifiers performance survey 2000-present. [Online]. Available: \url{https://ideas.ethz.ch/research/surveys/pa-survey.html}
\BIBentrySTDinterwordspacing

\bibitem{PAlimits}
J.~F. Buckwalter \emph{et~al.}, ``{Fundamental Limits of High-Efficiency Silicon and Compound Semiconductor Power Amplifiers in 100-300 GHz Bands},'' \emph{ITU J. on Future and Evolving Technol.}, vol. 2 (2021), no. 7 - Terahertz Commun., pp. 39--50, Oct. 2021.

\end{thebibliography}

\end{document}